\documentstyle[epsf,aps,psfig,12pt]{revtex}
\newcommand{\fslash}[1]{\mbox{$\!\not\!#1$}} 
\begin{document} 

\begin{center}
{\Large \bf Charge Neutrality Effects on 2-flavor Color Superconductivity} \\[3mm]
Mei Huang$^{1}$, Pengfei Zhuang$^{1}$, Weiqin Chao$^{2}$ \\[2mm]
$^1$ Physics Department, Tsinghua University, Beijing 100084, China \\
huangmei@mail.tsinghua.edu.cn \\
zhuangpf@mail.tsinghua.edu.cn \\[2mm]
$^2$ CCAST, Beijing 100080, China \\ 
Institute of High Energy Physics, \\
Chinese Academy of Sciences, Beijing 100039, China \\
chaowq@hp.ccast.ac.cn
\end{center}

\begin{abstract}

The effect of electric and color charge neutrality on 2-flavor color superconductivity
has been investigated. It has been found that the effect of the charge neutrality
on a 2-flavor quark system is very different from that on a 3-flavor
system. The BCS diquark pair in 2-flavor color superconducting phase has been largely
reduced by the large Fermi momentum of electron,
the diquark condensate first increases with the increase of
quark's chemical potential $\mu$, then decreases rapidly and finally
disappears at about $\mu=535 MeV$, at which the thermodynamic potential 
equals to that in the neutral normal 2-flavor quark matter.

\end{abstract}

\newpage

\section{Introduction}

Because of the existence of an attractive interaction in the anti-triplet 
quark-quark channel in QCD, the cold and dense quark matter has been 
believed to favor the formation of diquark
condensate and being in the superconducting phase \cite{love}-\cite{raja2}. 

The QCD phase structure at high baryon density is determined by how many kinds of
quarks can participate in the pairing. If we have an ideal system with two massless 
quarks $u$ and $d$, and the three colored  $u$ quarks and three colored $d$ quarks 
have the same Fermi surface, the system will be in 2 flavor color superconducting 
phase (2SC); and if we have an ideal system with three massless quarks $u$, $d$ and $s$,
and all the nine quarks have the same Fermi surface, then all the
quarks will participate in pairing and form a color-flavor-locking phase (CFL) \cite{cfl}. 

In the real world, the strange quark mass $m_s$ should be considered, 
which will reduce the strange quark's Fermi surface. If $m_s$ is large enough, 
there will be no pairing between $us$ and $ds$, and a 2SC+s phase 
will be favored \cite{unlock}. 

Also, in reality, the system should be electric and color neutral, the existence of
eletrons in the system will shift the Fermi surface of the two pairing quarks, and
the system can be in BCS phase or crystalline phase \cite{crystal} or normal quark 
matter, depending on how large the difference in the chemical potentials 
of the two pairing quarks is.

In a recent paper \cite{absence2sc}, it has been argued that the 2SC+s phase would 
not appear if the electric and color charge neutrality condition is added, which is also 
agreed with the results in \cite{absence2} by using SU(3) NJL model and taking into 
account the self-consistent effective quark mass $m_f(\mu)$, where $f$ refers to $u,d,s$.
It is found in \cite{absence2} that the charge neutrality has a large effect on the
$s$ quark mass, i.e., $m_s(\mu)$ begins to decrease at a smaller $\mu \simeq 400 MeV$ 
than the case if no charge neutrality is considered like in \cite{su3njl1} \cite{su3njl2}, 
where $m_s(\mu)$ begins to decrease at about $\mu \simeq 550 MeV$ . 

In this paper, complementary to \cite{absence2}, we investigate the effect of
charge neutrality on 2-flavor color superconductivity assuming that 
there is no strange quark involved in the chemical potential regime 
$\mu< 550 MeV$.  

This paper is organized as following: in Sec. II, we extend our method in \cite{mei}
to derive the thermodynamic potential when charge neutrality is 
considered; the gap equations and charge neutrality conditions will be derived in Sec. III, 
and in Sec. IV, we will give our numerical results; the conclusions 
and discussions will be given at the end.

\section{Thermodynamic Potential}

\subsection{The Lagrangian}

Assuming that the strange quark does not appear in the system,
we use the SU(2) Nambu-Jona-Lasinio model, and  
only consider scalar, pseudoscalar mesons and 
scalar diquark. The Lagrangian density has the form as
\begin{eqnarray}
\label{lagr}
{\cal L} = {\bar q}(i\gamma^{\mu}\partial_{\mu}-m_0)q + 
   G_S[({\bar q}q)^2 + ({\bar q}i\gamma_5{\bf {\vec \tau}}q)^2 ]
 +G_D[(i {\bar q}^C  \varepsilon  \epsilon^{b} \gamma_5 q )
   (i {\bar q} \varepsilon \epsilon^{b} \gamma_5 q^C)],
\end{eqnarray}
where $q^C=C {\bar q}^T$, ${\bar q}^C=q^T C$ 
are charge-conjugate spinors, $C=i \gamma^2 \gamma^0$ is the charge 
conjugation matrix (the superscript $T$ denotes the transposition operation),
the quark field $q \equiv q_{i\alpha}$ 
with $i=1,2$ and $\alpha=1,2,3$ is a flavor 
doublet and color triplet, as well as a four-component Dirac spinor, 
${\vec \tau}=(\tau^1,\tau^2,\tau^3)$ are Pauli matrices in the flavor 
space, and $(\varepsilon)^{ik} \equiv \varepsilon^{ik}$,
$(\epsilon^b)^{\alpha \beta} \equiv \epsilon^{\alpha \beta b}$ are  
antisymmetric tensors in the flavor and color spaces respectively.  
 $G_S$ and $G_D$ are independent effective coupling constants in the 
scalar quark-antiquark and scalar diquark channel. 

After bosonization, one can obtain the linearized version of the model (\ref{lagr}) 
\begin{eqnarray}
\label{lagr2}
\tilde{\cal L} & =  & {\bar q}(i\gamma^{\mu}\partial_{\mu}-m)q - 
    \frac{1}{2}\Delta^{*b} (i{\bar q}^C  \varepsilon \epsilon^{b}\gamma_5 q )
  -\frac{1}{2}\Delta^b (i {\bar q}  \varepsilon  \epsilon^{b} \gamma_5 q^C) \nonumber \\
  & & -\frac{\sigma^2}{4G_S}
  -\frac{\Delta^{*b}\Delta^{b}}{4G_D},
\end{eqnarray}
where we have assumed that there will be no pion condensate
and introduced the constituent quark mass 
\begin{eqnarray}
m=m_0+ \sigma.
\end{eqnarray}

From general considerations, there should be eight scalar diquark 
condensates \cite{rischke1}\cite{rischke3}. In the case of the NJL type model, 
the diquark condensates related to momentum vanish, and there
is only one $ 0^+$ diquark gap with Dirac structure 
$\Gamma=\gamma_5$ for massless quark, and another $0^+$ 
diquark condensate with Dirac structure
$\Gamma=\gamma_0\gamma_5$ at nonzero quark mass.
In this paper, we assume that the contribution of the diquark condensate with 
$\Gamma=\gamma_0\gamma_5$ is small, and only consider the diquark condensate
with $\Gamma=\gamma_5$. The diquark condensate with Dirac 
structure $\gamma_5\gamma_0$ has been recently discussed in \cite{todd}.

We can choose the diquark condenses in the third color direction,
i.e., only the first two colors participate in the condensate, while 
the third one does not. 

The model is non-renormalizable, and a momentum cut-off $\Lambda$ should be
introduced. The parameters $G_S$ and $\Lambda$ in the chiral limit $m_0=0$
 are fixed as
\begin{eqnarray}
G_S=5.0163 {\rm GeV}^{-2}, \Lambda=0.6533 {\rm GeV}.
\end{eqnarray}
The corresponding effective mass $m=0.314 {\rm GeV}$, and we will choose 
$G_D=3/4 G_S$ in our numerical calculations. 

%%%%%%%%%%
\subsection{Partition function and thermodynamic potential}
%%%%%%%%%%
%%%%%%%%%%

The partition function of the grand canonical ensemble can be evaluated 
by using standard method \cite{kapusta} \cite{bellac},
\begin{eqnarray}
\label{part}
{\cal Z}=N' \int [d {\bar q} ][d q] exp\{ \int_0^{\beta} d \tau \int d^3{\vec x} 
  ~ ( \tilde{\cal L} +\mu {\bar q} \gamma_0 q)\},
\end{eqnarray}
where $\beta=1/T$ is the inverse of temperature $T$, and 
$\mu$ is the chemical potential. When electric and color charge 
neutrality is considered, the chemical
potential $\mu$ is a diagonal $6 \times 6$ matrix in flavor and color space,
and can be expressed as
\begin{eqnarray}
\mu=diag(\mu_1,\mu_2,\mu_3,\mu_4,\mu_5,\mu_6),
\end{eqnarray}
where $\mu_1, \mu_2, \mu_3$ are for the three colored $u$ quarks,
and $\mu_4, \mu_5, \mu_6$ are for the three colored $d$ quarks.

Like in \cite{absence2sc} \cite{absence2}, the chemical potential
for each color and flavor quark is specified by its electric and color charges
$\mu_i=\mu-Q_e \mu_e +T_3 \mu_{3c}+T_8 \mu_{8c}$, where $Q_e,
T_3$ and $T_8$ are generators of $U(1)_Q, U(1)_3$ and $U(1)_8$.
Because the diquark condenses in the third color direction, and the first
two colored quarks are degenerate, we can assume $\mu_{3c}=0$. 
For the same flavor, the difference of chemical potentials between the first 
two colored quarks and the third colored quark is induced by $\mu_{8c}$,
and for the same color, the difference of chemical potentials between $u$ and $d$
is induced by $\mu_e$.

The explicit expressions for each color and flavor quark's chemical potential are:
\begin{eqnarray}
\mu_1=\mu_2=\mu-\frac{2}{3}\mu_e+\frac{1}{3}\mu_{8c}, \nonumber \\
\mu_4=\mu_5=\mu+\frac{1}{3}\mu_e+\frac{1}{3}\mu_{8c}, \nonumber \\
\mu_3=\mu-\frac{2}{3}\mu_e-\frac{2}{3}\mu_{8c}, \nonumber \\
\mu_6=\mu+\frac{1}{3}\mu_e-\frac{2}{3}\mu_{8c}.
\end{eqnarray}

For the convenience of calculations, we define 
the mean chemical potential ${\bar \mu}$ for the pairing quarks 
$q_1, q_5$, and $q_2, q_4$
\begin{eqnarray}
{\bar \mu}=\frac{\mu_1+\mu_5}{2}=\frac{\mu_2+\mu_4}{2}
=\mu-\frac{1}{6}\mu_e+\frac{1}{3}\mu_{8c},
\end{eqnarray}
and the difference of the chemical potential $\delta\mu$  
\begin{eqnarray}
\delta\mu = \frac{\mu_5-\mu_1}{2}=\frac{\mu_4-\mu_2}{2}=\mu_e /2.
\end{eqnarray}

Because the third colored $u$ and $d$, i.e., the 3rd and 6th quarks
do not participate in the diquark condensate, the partition function 
can be written as a product of three parts, 
\begin{eqnarray}
\label{part}
{\cal Z}={\cal Z}_{const} {\cal Z}_{36} {\cal Z}_{15,24} .
\end{eqnarray}
The constant part is
\begin{eqnarray}
{\cal Z}_{const}=N'{\rm exp} \{- \int_0^{\beta} d \tau \int d^3{\vec x} ~ 
[\frac{\sigma^2}{4 G_S}
      +\frac{\Delta^{*}\Delta}{4 G_D} \}.
\end{eqnarray}

${\cal Z}_{36}$ part is for the unpairing quarks $q_3$($u_3$) and $q_6$($d_3$), and 
${\cal Z}_{15,24}$ part is for the quarks participating in pairing, $q_1$ ($u_1$) paired 
with $q_5$ ($d_2$), and $q_2$($u_2$) paired with $q_4$ ($d_1$). In the following two subsections,
we will derive the contributions of ${\cal Z}_{36}$ and ${\cal Z}_{15,24}$.

\subsubsection{Calculation of ${\cal Z}_{36}$}

Introducing the  8-spinors for $q_3$ and $q_6$,
\begin{equation}
{\bar \Psi_{36}} = ( {\bar q}_3, {\bar q}_6; {\bar q}_3^C,{\bar q}_6^C ),
\end{equation}
we can express ${\cal Z}_{36}$ as
\begin{eqnarray}
\label{zq3}
 {\cal Z}_{36} & =  & \int[d \Psi_{36}]{\rm exp}\{\frac{1}{2}\sum_{n,{\vec p}}  
 ~ {\bar \Psi}_{36}\frac{[G_0^ {-1}]_{36}}{T}\Psi_{36} \} \nonumber \\
& = & {\rm Det} ^{1/2}(\beta [G_0^{-1}]_{36}),
\end{eqnarray}
where the determinantal operation ${\rm Det}$ is to be carried out over the Dirac, color, 
flavor and the momentum-frequency space, and $[G_0^{-1}]_{36}$ has the form of
\begin{equation}
[G_0^{-1}]_{36} = 
    \left( 
          \begin{array}{cccc}
             [G_0^{+}]_{3}^{-1}  &  0 & 0 & 0 \\  
            0 &   [G_0^{+}]_{6}^{-1}  & 0 & 0 \\
             0 &  0 &  [G_0^{-}]_{3}^{-1}  & 0  \\
              0 &  0 & 0 &  [G_0^{-}]_{6}^{-1} 
              \end{array}  
             \right),
\end{equation}
with 
\begin{eqnarray}
[G_0^{\pm}]^{-1}_{i}=  {\fslash p} \pm {\fslash \mu_{i}} -m.
\end{eqnarray}
Here we have used ${\fslash p}=p_{\mu}\gamma^{\mu}$ and ${\fslash \mu_{i}} = \mu_i \gamma_0$.

For the two quarks not participating in the diquark condensate, 
from Eq. (\ref{zq3}), we can have 
\begin{eqnarray}
{\rm ln} {\cal Z}_{36} & = & \frac{1}{2} {\rm ln}  {\rm Det}
(\beta [G_0^{-1}]_{36}) \nonumber \\
 & = & \frac{1}{2} {\rm ln}[{\rm Det} (\beta [G_0^+]_3^{-1}) 
 {\rm Det} (\beta [G_0^-]_3^{-1})] [{\rm Det} (\beta [G_0^+]_6^{-1}) 
 {\rm Det} (\beta [G_0^-]_6^{-1})].
\end{eqnarray}
We first work out 
\begin{eqnarray}
 & [{\rm Det} (\beta [G_0^+]_3^{-1})  {\rm Det} (\beta [G_0^-]_3^{-1}) ] & =  \beta^4
  [p_0^2-E_{3}^{+^2}] [p_0^2-E_{3}^{-^2}],  \nonumber \\
 &  [{\rm Det} (\beta [G_0^+]_6^{-1}) {\rm Det} (\beta [G_0^-]_6^{-1})] & =  \beta^4 
 [p_0^2-E_6^{+^2}][p_0^2-E_6^{-^2}], 
\end{eqnarray}
with $E_3^{\pm}=E \pm \mu_3$ and $ E_6^{\pm}=E \pm \mu_6$ 
where $E=\sqrt{{\bf p}^2+m^2}$.
Considering the determinant in the flavor, color, spin spaces and momentum-frequency 
space, we get the expression 
\begin{eqnarray}
{\rm ln} {\cal Z}_{36}   & = &  
   \sum_n \sum_{\vec p} \{ {\rm ln} ( \beta^2 [ p_0^2- E_{3}^{+^2} ] ) 
   +  {\rm ln} ( \beta^2  [ p_0^2- E_3^{-^2} ] )  \nonumber \\
   & &  + {\rm ln} ( \beta^2 [ p_0^2- E_6^{+^2} ] ) 
   +  {\rm ln} ( \beta^2  [ p_0^2- E_6^{-^2} ] ) \}.
\end{eqnarray}

\subsubsection{Calculation of ${\cal Z}_{15,24}$}

The calculation of ${\cal Z}_{15,24}$ here is much more complicated than that
when the two pairing quarks have the same Fermi surface \cite{mei}.

Also, we introduce the Nambu-Gokov formalism for $q_1, q_2, q_4$ and $q_5$, i.e.,
\begin{equation}
{\bar \Psi} = ( {\bar q}_1, {\bar q}_2, {\bar q}_4, {\bar q}_5; {\bar q}_1^C,
{\bar q}_2^C,{\bar q}_4^C,{\bar q}_5^C ).
\end{equation} 
The ${\cal Z}_{15,24}$ can have the simple form as
\begin{eqnarray}
\label{zq12}
{\cal Z}_{15,24} & = & \int[d \Psi]{\rm exp} \{\frac{1}{2}  \sum_{n,{\vec p}} ~
{\bar \Psi}\frac{G^{-1}_{15,24}}{T}  \Psi \}  \nonumber \\
& = & {\rm Det}^{1/2}(\beta [G^{-1}]_{15,24}),
\end{eqnarray}
where
\begin{equation}
{\rm G}^{-1}_{15,24} = 
    \left( 
          \begin{array}{cc}
             [G_0^{+}]_{15,24}^{-1}  &  \Delta^{-} \\  
            \Delta^{+}  &  [ G_0^{-}]_{15,24}^{-1} 
              \end{array}  
             \right),
\end{equation}
with
\begin{eqnarray}
[G_0^{\pm}]_{15,24}^{-1}=
\left( 
          \begin{array}{cccc}
            [G_0^{\pm}]_{1}^{-1}  &  0  & 0 & 0\\  
            0 &  [G_0^{\pm}]_{2}^{-1}  & 0 & 0\\
            0  & 0 & [G_0^{\pm}]_{4}^{-1}  &  0 \\  
            0 & 0  & 0 &  [G_0^{\pm}]_{5}^{-1}
              \end{array}  
             \right),
\end{eqnarray}
and the matrix form for $\Delta^{\pm}$  is
\begin{eqnarray} 
\Delta^{-} = -i \Delta \gamma_5 \left( 
          \begin{array}{cccc}
            0  &  0  & 0 & 1\\  
            0 & 0 & -1 & 0\\
            0  & -1 & 0  &  0 \\  
            1 & 0  & 0 & 0 
              \end{array}  
             \right),  \  \ 
\Delta^{+} = \gamma^0 (\Delta^{-})^{\dagger} \gamma^0.
\end{eqnarray}

From Eq. (\ref{zq12}), we have 
\begin{eqnarray}
\label{detg}
{\rm ln} Z_{15,24}=\frac{1}{2} {\rm ln} {\rm Det} (\beta G^{-1}_{15,24}).
\end{eqnarray}

For a $2 \times 2$ matrix with elements $A,B,C$ and $D$, we have the identity
\begin{eqnarray}
\label{identity}
{\rm Det} \left(  \begin{array}{cc}
            A  &  B  \\  
           C & D
              \end{array}  
             \right) = {\rm Det} ( -CB + CAC^{-1}D ) ={\rm Det}(-BC+BDB^{-1}A).
\end{eqnarray}

Replacing $A,B,C$ and $D$ with the corresponding elements of $G_{15,24}^{-1}$, 
we have 
\begin{eqnarray}
{\rm Det} ({\rm G}_{15,24}^{-1}) 
& = &{\rm Det} D_1  =  {\rm Det} (- \Delta^{+} \Delta^{-} 
+ \Delta^{+}[G_0^{+}]_{15,24}^{-1} 
[\Delta^{-}]^{-1}[G_0^{-}]_{15,24}^{-1} ) \nonumber \\
& = & {\rm Det} D_2 = {\rm Det} ( - \Delta^{-} \Delta^{+} 
+ \Delta^{-} [G_0^{-}]_{15,24}^{-1} [\Delta^{+}]^{-1}  [G_0^{+}]_{15,24}^{-1} ).
\end{eqnarray}
Using the massive energy projectors $\Lambda_{\pm}$ in \cite{mei}
for each flavor and color quark, we can work out the diagocal matrix 
$D_1$ and $D_2$ as
\begin{eqnarray}
(D_1)_{11}= (D_1)_{22} = [(p_0 + \delta \mu)^2 - [{\bar E}_{\Delta}^-]^2] \Lambda_{-}
+ [(p_0 + \delta \mu)^2 - [{\bar E}_{\Delta}^+]^2] \Lambda_{+} \nonumber \\
(D_1)_{33}= (D_1)_{44} = [(p_0 - \delta \mu)^2 - [{\bar E}_{\Delta}^-]^2] \Lambda_{-}
+ [(p_0 - \delta \mu)^2 - [{\bar E}_{\Delta}^+]^2] \Lambda_{+} \nonumber \\
(D_2)_{11}= (D_2)_{22} = [(p_0 - \delta \mu)^2 - [{\bar E}_{\Delta}^+]^2] \Lambda_{-}
+ [(p_0 - \delta \mu)^2 - [ {\bar E}_{\Delta}^-]^2] \Lambda_{+} \nonumber \\ 
(D_2)_{33}= (D_2)_{44} = [(p_0 + \delta \mu)^2 - [{\bar E}_{\Delta}^+]^2] \Lambda_{-}
+ [(p_0 + \delta \mu)^2 - [{\bar E}_{\Delta}^-]^2] \Lambda_{+}, 
\end{eqnarray}
where ${\bar E}_{\Delta}^{\pm} = \sqrt{ (E \pm {\bar \mu})^2+\Delta^2}$.

With the above equations, Eq. (\ref{detg}) can be expressed as
\begin{eqnarray}
{\rm ln} {\cal Z}_{15,24} & = &  
2  \sum_n \sum_p \{ {\rm ln}[\beta^2 (p_0^2-({\bar E}_{\Delta}^-+\delta\mu))^2]
 + {\rm ln}[\beta^2 (p_0^2-({\bar E}_{\Delta}^- - \delta\mu))^2 ] \nonumber \\
  & + & {\rm ln}[\beta^2 (p_0^2-({\bar E}_{\Delta}^+ + \delta\mu))^2 ]
 + {\rm ln}[\beta^2 (p_0^2-({\bar E}_{\Delta}^+ -\delta\mu))^2] \}.
\end{eqnarray}

%%%%%%%%%%%%%%%%%%%%
\subsection{The thermodynamic potential}
%%%%%%%%%%%%%%%%%%%%%

Using the helpful relation 
\begin{eqnarray}
\label{lnzf}
{\rm ln}{\cal Z}_f  & = &  \sum_n{\rm ln}[\beta^2(p_0^2-E_p^2)] \nonumber \\
& = &\beta [E_p +2T {\rm ln}(1+e^{-\beta E_p})],
\end{eqnarray}
we can evaluate the thermodynamic potential of the quark system
\begin{eqnarray}
\label{potential}
\Omega_q & = &\frac{m^2}{4G_S}+\frac{\Delta^2}{4G_D} 
- 2 \int\frac{d^3 p}{(2\pi)^3} [ 2 E +  T{\rm ln}(1+exp(-\beta E_3^+)) \nonumber \\
 & &  + T {\rm ln}(1+ exp(-\beta E_3^-))   +  T{\rm ln}(1+exp(-\beta E_6^+)) 
+ T {\rm ln}(1+ exp(-\beta E_6^-)) \nonumber \\
& &  + 2 {\bar E}_{\Delta}^+  +
 2 {\bar E}_{\Delta}^-  +2T{\rm ln}(1+exp(-\beta {\bar E}_{\Delta^+}^{+}))   
+2T{\rm ln}(1+exp(-\beta {\bar E}_{\Delta^-}^{+})) \nonumber \\
& & +2T{\rm ln}(1+exp(-\beta {\bar E}_{\Delta^+}^{-}))   
+2T{\rm ln}(1+exp(-\beta {\bar E}_{\Delta^-}^{-})) ]
\end{eqnarray}
with $ {\bar E}_{\Delta^\pm}^{\pm}= {\bar E}_{\Delta}^{\pm} \pm \delta \mu $.

For the total thermodynamic potential, we should include the contribution from
the electron gas, $\Omega_e$. Assuming the electron's mass is zero,
we have
\begin{eqnarray}
\Omega_e=-\frac{\mu_e^4}{12 \pi^2}.
\end{eqnarray}

The total thermodynamic potential of the system is  
\begin{eqnarray}
\Omega=\Omega_q+\Omega_e.
\label{omega}
\end{eqnarray}

\section{Gap Equations and Charge Neutrality Condition}

From the thermodynamic potential Eq.(\ref{omega}), we can
derive the gap equations of the order parameters $m$ and  $\Delta$ 
for the chiral and color superconducting phase transitions. 

\subsection{Gap equation for quark mass}

The gap equation for quark mass can be derived by using
\begin{eqnarray}
\frac{\partial \Omega}{\partial m}=0.
\end{eqnarray}
The explicit expression for the above equation is 
\begin{eqnarray}
 m [1- 4G_S &  & \int\frac{d^3{\bf p}}{(2\pi)^3}\frac{1}{E} 
[ 2\frac{{\bar E}^-}{{\bar E}_{\Delta}^-}(1- {\tilde f}({\bar E}_{\Delta^+}^-) 
- {\tilde f}({\bar E}_{\Delta^-}^- ) ) \nonumber \\
&  &  + 2 \frac{{\bar E}^+}{{\bar E}_{\Delta}^+}(1- {\tilde f}({\bar E}_{\Delta^+}^+)
- {\tilde f}({\bar E}_{\Delta^-}^+ ) )  \nonumber \\
&  & + 
(2- {\tilde f}(E_3^+) - {\tilde f}(E_3^-) - {\tilde f}(E_6^+) - {\tilde f}(E_6^-) )] =0,
\end{eqnarray}
with the Fermi distribution function ${\tilde f}(x)=1/(exp\{\beta x\} + 1)$.
$m=0$ corresponds to the chiral symmetric phase, $m \neq 0$ corresponds to
the chiral symmetry breaking phase.

\subsection{Gap equation for diquark condensate}
Using 
\begin{eqnarray}
\frac{\partial \Omega}{\partial \Delta}=0,
\end{eqnarray}
we can derive the gap equation for diquark condensate
\begin{eqnarray}
 \Delta [1- & & 4G_D\int\frac{d^3{\bf p}}{(2\pi)^3} 
[ 2 \frac{1}{{\bar E}_{\Delta}^-}(1- {\tilde f}({\bar E}_{\Delta^+}^-) 
- {\tilde f}({\bar E}_{\Delta^-}^- ) ) \nonumber \\
&  & + 2 \frac{1}{{\bar E}_{\Delta}^+}(1- {\tilde f}({\bar E}_{\Delta^+}^+)
- {\tilde f}({\bar E}_{\Delta^-}^+ ) ) ] =0.
\end{eqnarray}

\subsection{Color charge neutrality}

The color charge neutrality condition is to choose $\mu_{8c}$ such that the
system has zero net charge $T_8$, i.e.,
\begin{eqnarray}
T_8=\frac{\partial \Omega}{\partial \mu_{8c}}=0.
\end{eqnarray}
Evaluating the above equation, we have the color charge neutrality condition as
\begin{eqnarray}
  \int\frac{d^3{\bf p}}{(2\pi)^3} &  &
[ - \frac{{\bar E}^-}{{\bar E}_{\Delta}^-}(1- {\tilde f}({\bar E}_{\Delta^+}^-) 
- {\tilde f}({\bar E}_{\Delta^-}^- ) ) \nonumber \\
&  & +  \frac{{\bar E}^+}{{\bar E}_{\Delta}^+}(1- {\tilde f}({\bar E}_{\Delta^+}^+)
- {\tilde f}({\bar E}_{\Delta^-}^+ ) ) \nonumber \\
& & +  ({\tilde f}(E_3^+) - {\tilde f}(E_3^-)) + ({\tilde f}(E_6^+) - {\tilde f}(E_6^-) )] =0.
\end{eqnarray}

\subsection{Electric charge neutrality}
Similarly, the electric charge neutrality condition is to choose $\mu_e$ such that the
system has zero net eletric charge $Q_e$, i.e.,
\begin{eqnarray}
Q_e=\frac{\partial \Omega}{\partial \mu_e}=0.
\end{eqnarray}
From the above equation, we obtain
\begin{eqnarray}
  &  & \int\frac{d^3{\bf p}}{(2\pi)^3} 
[ 2 \frac{{\bar E}^-}{{\bar E}_{\Delta}^-}(1- {\tilde f}({\bar E}_{\Delta^+}^-) 
- {\tilde f}({\bar E}_{\Delta^-}^- ) ) \nonumber \\
&  & - 2 \frac{{\bar E}^+}{{\bar E}_{\Delta}^+}(1- {\tilde f}({\bar E}_{\Delta^+}^+)
- {\tilde f}({\bar E}_{\Delta^-}^+ ) ) \nonumber \\
& & + 6 [{\tilde f}({\bar E}_{\Delta^+}^-) + {\tilde f}({\bar E}_{\Delta^+}^+)
 - {\tilde f}({\bar E}_{\Delta^-}^+) - {\tilde f}({\bar E}_{\Delta^-}^-) \nonumber \\
& & + 4 ({\tilde f}(E_3^+) - {\tilde f}(E_3^-)) - 2 ({\tilde f}(E_6^+) - {\tilde f}(E_6^-) )] +
\frac{\mu_e^3}{\pi^2} = 0.
\end{eqnarray}

\section{Numerical Results}

In the numerical results, we investigate the following four different cases:
1), no color superconducting phase, no charge neutrality condition;
2), no color superconducting phase, electirc charge neutrality condition added, the electron's
chemical potential in this phase will be characterized by $\mu_e^0$;
3), color superconducting phase exists, no charge neutrality condition,  the diquark condensate
in this phase will be characterized by $\Delta_0$; 
4), color superconducting phase exists, electric and color charge neutrality condition added, the
electron's chemical potential and diquark condensate in this phase will be characterized by
$\mu_e$ and $\Delta$ respectively. 

\begin{figure}[ht]
\centerline{\epsfxsize=16cm\epsffile{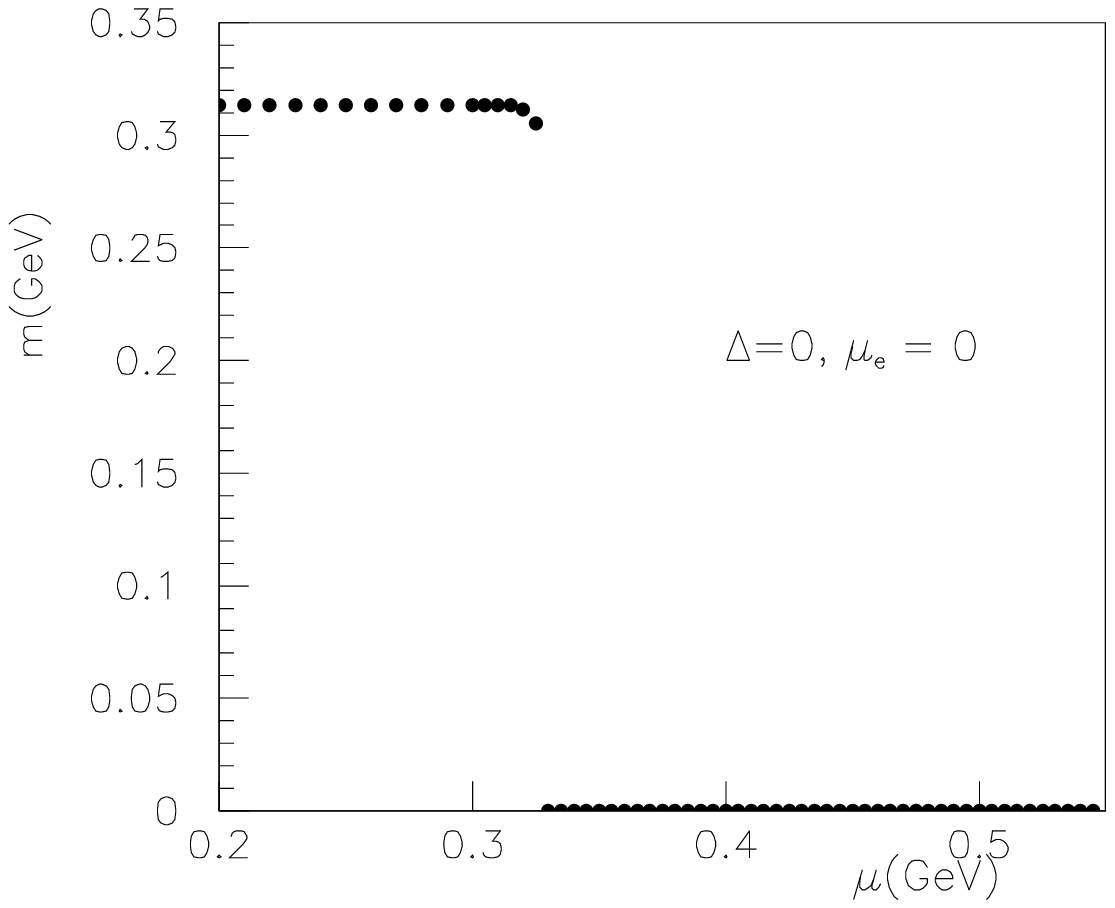}}
\caption{The quark mass $m$ as a function of chemical potential $\mu$ in the case of 
no color superconducting phase and no charge neutrality condition.}
\label{nodnoe_fig}
\end{figure}

Fig. \ref{nodnoe_fig} is for the chiral phase transition in the case of no electric neutrality 
considered, which is familiar to all of us, the quark posses a large constituent mass in the
low baryon density regime due to chiral symmetry breaking, and restores chiral symmetry
at high baryon density. 

\begin{figure}[ht]
\centerline{\epsfxsize=16cm\epsffile{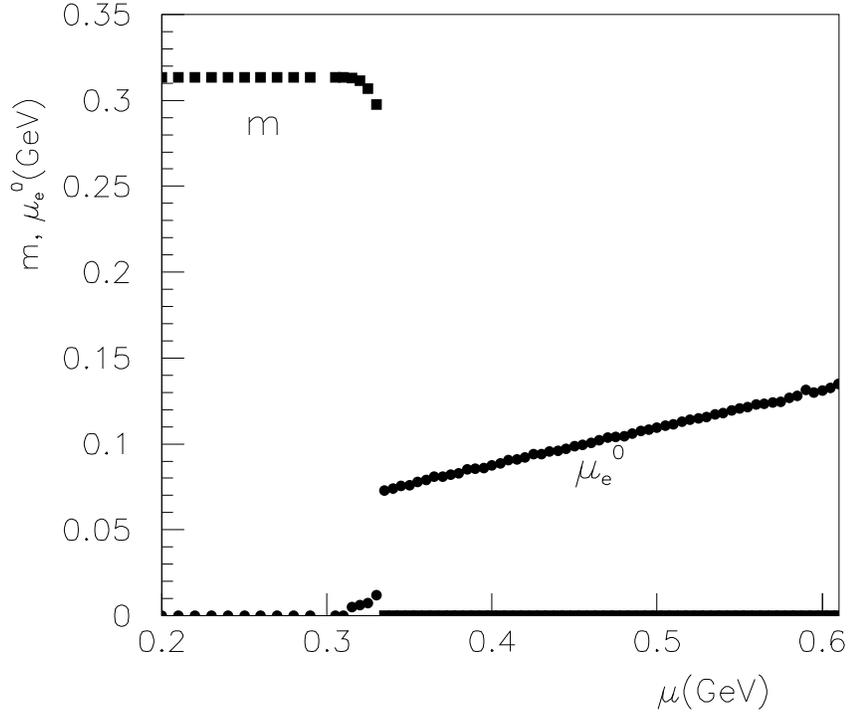}}
\caption{The quark mass $m$ (squares) and the electron's chemical potential 
$\mu_e^0$ (circles) as functions of chemical potential $\mu$ in the case of
neutral normal quark matter. }
\label{nodmue}
\end{figure}

Fig. \ref{nodmue} is the chiral phase diagram when the electric
charge neutrality is considered, the solid squares and the solid circles are for quark mass $m$
and electron's chemical potential $\mu_e^0$ respectively. It is found that, in the chiral symmetric
phase, there is a large Fermi surface for electrons, and the electron's chemical potential
$\mu_e^0$ increases linearly with the increase of quark's chemical potential $\mu$.

\begin{figure}[ht]
\centerline{\epsfxsize=16cm\epsffile{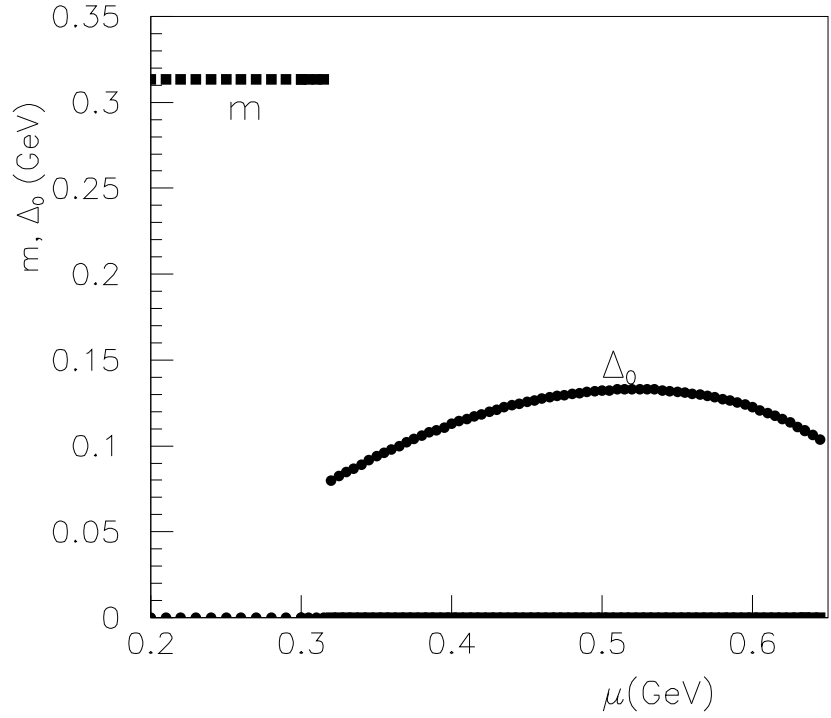}}
\caption{The quark mass $m$ (squares) and the diquark condensate $\Delta_0$ (circles)
as functions of chemical potential $\mu$ in the case of no charge neutrality
on color superconducting phase.}
\label{ddnoe}
\end{figure}

Fig. \ref{ddnoe} is the phase diagram for color superconductivity without charge neutrality
constraints, the quark mass $m$ (squares) and diquark condensate $\Delta_0$ (circles)
are plotted as functions of quark's chemical potential $\mu$. It is found that in the color
superconducting phase,  the magnitude of the diquark
condensate $\Delta_0$ is about $100 MeV$ by using the parameter $G_D=3/4 G_S$.
$\Delta_0$ first increases with increasing $\mu$, and this tendency stops at 
about $ \mu=530 MeV$, which is related to the momentum cut-off $\Lambda$.

\begin{figure}[ht]
\centerline{\epsfxsize=16cm\epsffile{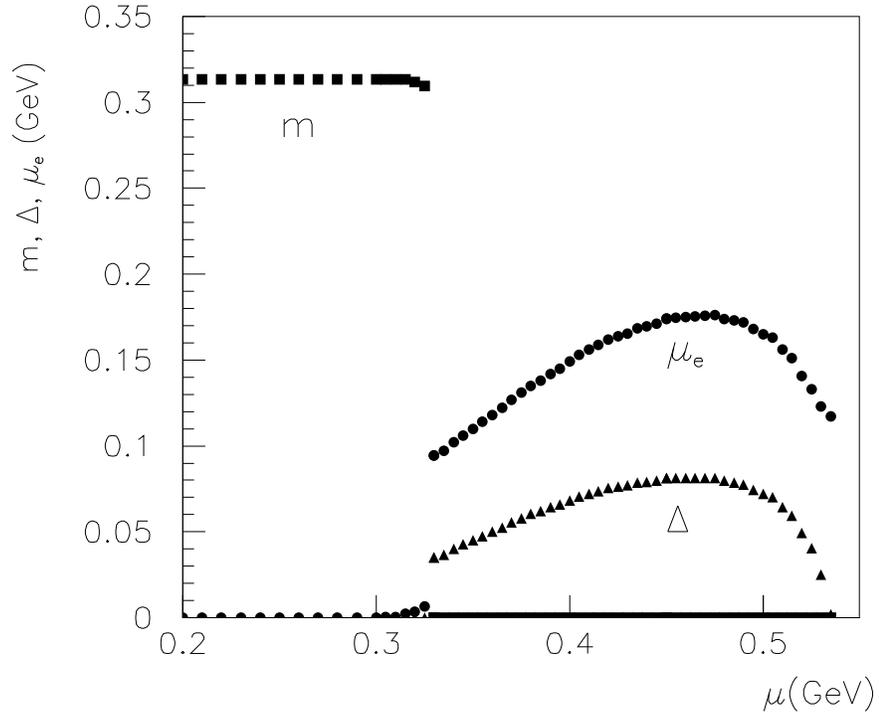}}
\caption{The quark mass $m$ (squares), the diquark condensate $\Delta$ (triangles)
and the electron's chemical potential $\mu_e$ (circles)
as functions of chemical potential $\mu$ in the case of considering charge neutrality
on color superconducting phase.}
\label{ddmue}
\end{figure}

Fig. \ref{ddmue} is the phase diagram for color superconductivity when electric and
color charge neutrality condition is considered,  the quark mass $m$ (squares), diquark 
condensate $\Delta$ (triangles) and the electron's chemical potential $\mu_e$ (circles)
are plotted as functions of quark's chemical potential $\mu$. It is found that when
chiral symmetry restores, the color superconducting phase appears. In the color
superconducting phase, it is found that both the diquark condensate $\Delta$ and
the electron's chemical potential first increase with increasing $\mu$,
and reaches their maximum at about $\mu=475 MeV$, then decrease rapidly with
increasing $\mu$, and the diquark condensate disappears at about $\mu=535 MeV$. 

\begin{figure}[ht]
\centerline{\epsfxsize=14cm\epsffile{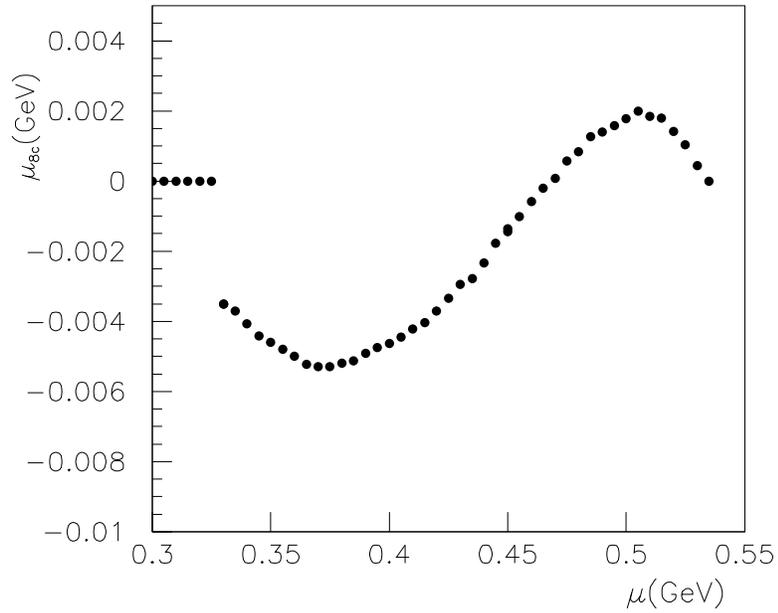}}
\caption{The $\mu_{8c}$ as a function of chemical potential $\mu$ 
in the case of considering charge neutrality on color superconducting phase.}
\label{mu8c}
\end{figure}

In Fig. \ref{mu8c}, we show the chemical potential 
$\mu_{8c}$ as a function of $\mu$, which ensures the system having zero 
net color charge $T_8$. It is found
that $\mu_{8c}$ can be negative and positive, but its value is very small, only about
several $MeV$. It means that for the same flavor quark $u$ or $d$, the difference of Fermi
momenta between the third colored quark and the first two colored quarks can be
neglected.

\begin{figure}[ht]
\centerline{\epsfxsize=16cm\epsffile{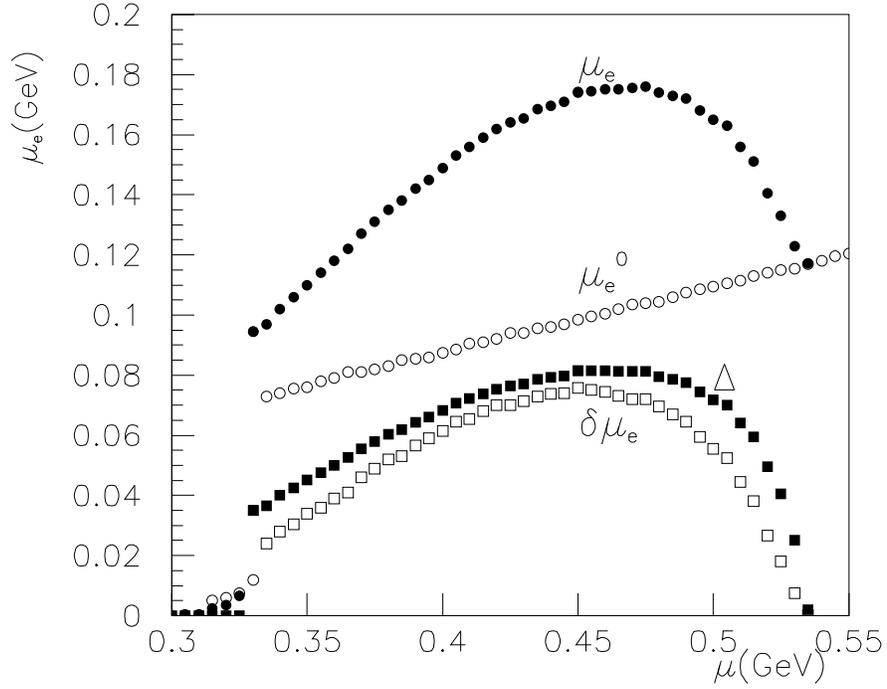}}
\caption{The electron's chemical potential in the neutral normal quark matter 
$\mu_e^0$ (light circles) and 
in the neutral superconducting phase $\mu_e$ (solid circles) as functions of 
chemical potential $\mu$, and $\delta\mu_e=\mu_e-\mu_e^0$ (light squares)
comparing with $\Delta$ (solid squares) as functions of $\mu$.  }
\label{mue}
\end{figure}

\begin{figure}[ht]
\centerline{\epsfxsize=16cm\epsffile{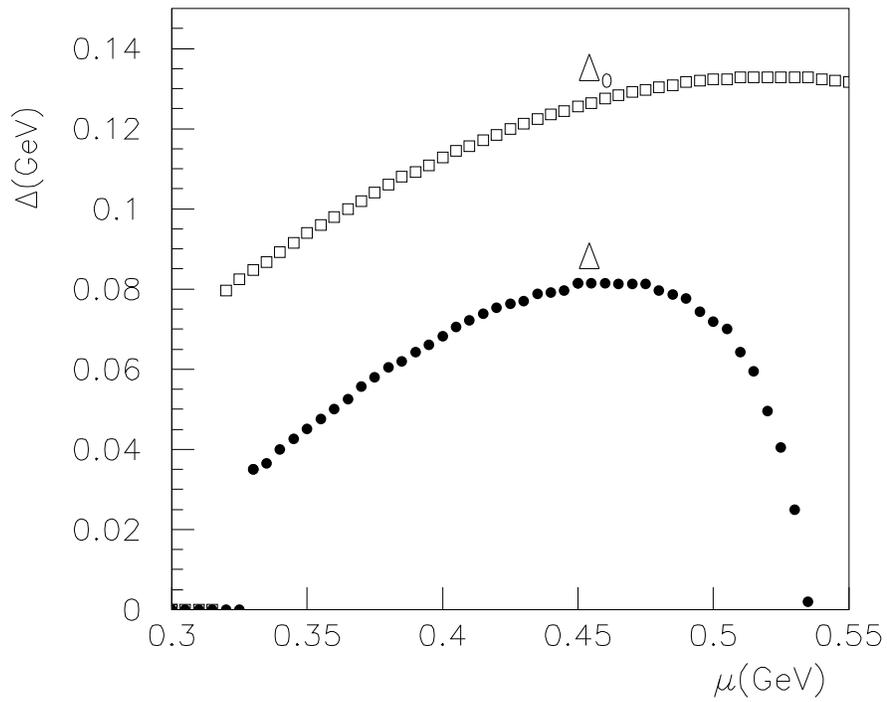}}
\caption{The diquark condensate 
in the electric charged $\Delta_0$ (light squares) and in the neutral 
$\Delta$ (solid circles) superconducting 
phase as a function of chemical potential $\mu$.}
\label{dd}
\end{figure}

In order to explicitly see what happens to a color superconducting phase when
charge neutrality condition is added, we compare the electron's chemical potential
in the neutral normal quark matter $\mu_e^0$ (light circles) and 
in the neutral superconducting
phase $\mu_e$ (solid circles) in Fig. \ref{mue}, and compare the diquark condensate 
$\Delta_0$ (light squares) and $\Delta$ (solid circles) in the electric charged  and
neutral superconducting phases in Fig. \ref{dd}.  

From Fig. \ref{mue}, we find that the electron has a larger Fermi
surface in the neutral superconducting phase than that in the neutral normal quark matter.
This is because in the superconducting phase, the isospin is mainly carried by unpaired
quarks only, and $\mu_e$ has to be larger in order to get the same isospin density. 
$\mu_e$ is equal to $\mu_e^0$ at about  $\mu=535 MeV$, at which the diquark condensate 
disappears. We seperate $\mu_e$ into
two parts as $\mu_e=\mu_e^0+\delta\mu_e$, where $\delta\mu_e$ reflects the effect induced by
diquark condensate, comparing $\delta\mu_e$ (light squares) and $\Delta$ (solid squares)
in Fig.  \ref{mue}, we can see that $\delta\mu_e$ increases when $\Delta$ increases,
and decreases when $\Delta$ decreases, and when $\Delta=0$, $\delta\mu_e=0$, too.

From Fig. \ref{dd}, we can see that the diquark condensate $\Delta$ in the 
neutral superconducting phase has a much smaller
value than $\Delta_0$ in the charged superconducting phase, it means that  
the difference of the Fermi surfaces of the two pairing quarks reduces the 
magnitude of the diquark condensate $\Delta$. 

From Fig. \ref{mue} and Fig. \ref{dd}, we can see a distinguished characteristic
in the neutral superconducting phase, i.e., both the electon's chemical potential $\mu_e$
and the magnitude of the diquark condensate $\Delta$ first increase with increasing $\mu$,
then at about $\mu=475 MeV$, decrease with increasing 
$\mu$. Now we try to understand this phenomena. 

The magnitude of the diquark condensate is not only affected by $\mu$, but also by 
$\delta\mu=\mu_e/2$, which describes the difference of Fermi surface between the
two paired quarks and reduces the diquark condensate as seen in Fig. \ref{dd}.
The diquark condensate first increases with increasing $\mu$, which is 
the result of the increasing density of states.
At about $\mu=475 MeV$,  where $\mu_d=\mu+1/3 \mu_e$ is about $530 MeV$,
the diquark condensate starts to be affected by the momentum cut-off $\Lambda$. Like that
in the charged color superconducting phase, the diquark condensate stops increasing with
$\mu$, so does $\mu_e$.
However, when $\mu> 475 MeV$, $\mu_e^0$ keeps increasing with $\mu$, 
which reduces the diquark condensate largely. This is why we see $\Delta$ decreases
with increasing $\mu$. Finally at a certain chemical potential $\mu=535 MeV$, 
where $\mu_d$ is about $580 MeV$, the diquark condenate disappears. 
Because $\delta\mu_e$ decreases with the decrease
of $\Delta$,  the total $\mu_e$ decreases in the chemical potential regime $\mu> 475 MeV$.

\begin{figure}[ht]
\centerline{\epsfxsize=14cm\epsffile{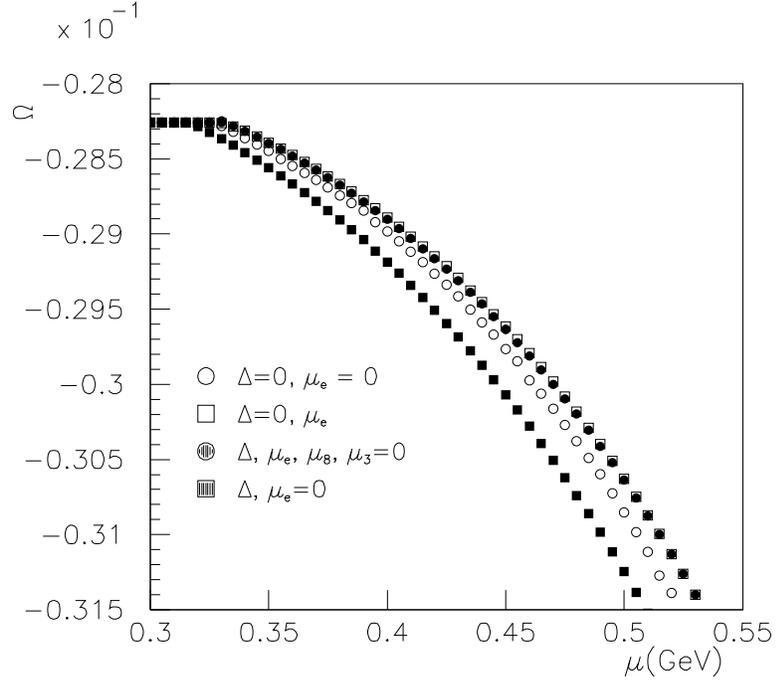}}
\caption{The thermodynamic potentials as functions of chemical potential $\mu$,
the light squares and circles are for the cases with and without charge neutrality and 
without diquark condensate, the dark circles and squares are for the cases with and 
without charge neutrality and with diquark condensate.  }
\label{pot}
\end{figure}

At last, in Fig.\ref{pot}, we show the thermodynamic potentials in the four different 
cases as functions of chemical potential $\mu$, the light squares and circles are
for the cases with and without charge neutrality and without diquark condensate,
the dark circles and squares are for the cases with and without charge neutrality 
and with diquark condensate. It can be seen that in the chiral symmetric phase,
the charged superconducting phase has the lowest thermodynamic potential,
and the charged normal quark matter has the second lowest $\Omega$.
The neutral superconducting phase has a little bit lower $\Omega$ than that
of the neutral quark matter in the chemical potential region $\mu < 535 MeV$,
and at $\mu=535 MeV$, the two $\Omega$s coincide with each other.
Therefore, in the chiral symmetric phase, the stable state of the neutral system 
is the color superconducting phase for $<\mu < 535 MeV$. At $\mu=535 MeV$,
the stable phase is the nomoral neutral quark matter.

\section{Conclusions}

We invetigated the effect of charge neutrality on a two flavor quark system.
It has been found tha the BCS diquark pair has been largely
reduced by the large Fermi momentum of electron,
the diquark condensate first increases with the increase of
quark's chemical potential $\mu$, then decreases rapidly and
diappears at about $\mu=535 MeV$, at which the thermodynamic potential 
equals to that in the neutral normal quark matter. In the chemical potential
region $330 MeV < \mu < 535 MeV$, the stable neutral system is in the color
superconducting phase if no strange quark involved. 

As we mentioned in the introduction, we did not consider the strange quark 
in the system, which should be considered and has been investigated 
self-consistently in \cite{absence2}. Comparing our
results with their results about the diquark gap $\Delta$, the electron's chemical
potential $\mu_e$ and the chemical potential $\mu_{8c}$ in 2SC, it can be 
found that the main difference lies in the chemical potential region $\mu > 400 MeV$.
In \cite{absence2}, the diquark condensate in the neutral superconducting phase
is largely reduced only in a small chemical potential region 
$370 MeV < \mu < 400 MeV$.
$\mu_e$ first increases with $\mu$ then
begins to decrease at about $\mu=400 MeV$.
As for $\mu_{8c}$, the tendency also becomes different from our results 
at about $\mu=400 MeV$.   
The reason lies in that,  in \cite{absence2}, the strange quark involves in the system 
when $\mu > 400 MeV$. This shows that the effect of the charge neutrality condition on
three- and two- flavor quark system is quite different.

%%%%%%%%%%
%%%%%%%%%%
\section*{Acknowledgements}
%%%%%%%%%%
%%%%%%%%%%

One of the authors (M.H.) thanks valuable discussions with Dr. T. Schafer.
This work was supported in part by China Postdoctoral Science Foundation, 
the NSFC under Grant Nos. 10105005, 10135030 and 19925519,  and the 
Major State Basic Research Development Program under Contract No. G2000077407.

\newpage

\newpage

\newpage

\newpage

\newpage

\newpage

\newpage

\newpage
\newpage

\end{document}